\documentclass[11pt,runningheads]{llncs}
\usepackage{amsmath,amsfonts}
\usepackage{algorithmic}
\usepackage{algorithm}
\usepackage{array}
\usepackage[caption=false,font=normalsize,labelfont=sf,textfont=sf]{subfig}
\usepackage{textcomp}
\usepackage{stfloats}
\usepackage{url}
\usepackage{verbatim}
\usepackage{graphicx}
\usepackage{cite}
\usepackage{multirow}
\usepackage{epsfig}
\usepackage[utf8]{inputenc}
\usepackage{array}
    \newcolumntype{P}[1]{>{\centering\arraybackslash}p{#1}}
    \newcolumntype{M}[1]{>{\centering\arraybackslash}m{#1}}

\usepackage{stackengine}

\def\BibTeX{{\rm B\kern-.05em{\sc i\kern-.025em b}\kern-.08em
    T\kern-.1667em\lower.7ex\hbox{E}\kern-.125emX}}
\usepackage{balance}

\usepackage{enumitem}
\usepackage{cuted}
\usepackage{cryptocode} 
\usepackage{fancybox,framed}
\usepackage[margin=1in]{geometry}

\usepackage{appendix}

\begin{document}
\title{Differentiated Security Architecture for Secure and Efficient Infotainment Data Communication in IoV Networks\thanks{This work was supported by Alibaba Group through Alibaba Innovative Research (AIR) Program and Alibaba-NTU Singapore Joint Research Institute (JRI), Nanyang Technological University, Singapore. This version of the contribution has been accepted for publication, after peer review (when applicable) but is not the Version of Record and does not reflect post-acceptance improvements, or any corrections. The Version of Record is available online at: https://doi.org/10.1007/978-3-031-23020-2\_16. Use of this Accepted Version is subject to the publisher’s Accepted Manuscript terms of use https://www.springernature.com/gp/open-research/policies/accepted-manuscript-terms.}}

\author{Jiani Fan\inst{1}\thanks{Corresponding Author} \and
Lwin Khin Shar\inst{2} \and
Jiale Guo\inst{1} \and
Wenzhuo Yang\inst{1} \and
Dusit Niyato\inst{1} \and
Kwok-Yan Lam\inst{1}
}

\authorrunning{Fan et al.}
\titlerunning{Differentiated Security Architecture for Infotainment Data Communication in IoV Network}
%
\institute{Nanyang Technological University, 50 Nanyang Ave, Singapore \\
\email{\{jiani001, jiale001, wenzhuo001,\}@e.ntu.edu.sg, \{DNIYATO, kwokyan.lam\}@ntu.edu.sg} \and
Singapore Management University, 81 Victoria St, Singapore \\
\email{lkshar@smu.edu.sg}}

\maketitle              
\begin{abstract}
This paper aims to provide differentiated security protection for infotainment data communication in Internet-of-Vehicle (IoV) networks. The IoV is a network of vehicles that uses various sensors, software, built-in hardware, and communication technologies to enable information exchange between pedestrians, cars, and urban infrastructure. Negligence on the security of infotainment data communication in IoV networks can unintentionally open an easy access point for social engineering attacks. The attacker can spread false information about traffic conditions, mislead drivers in their directions, and interfere with traffic management. Such attacks can also cause distractions to the driver, which has a potential implication for the safety of driving. 
The existing literature on IoV communication and network security focuses mainly on generic solutions. In a heterogeneous communication network where different types of communication coexist, we can improve the efficiency of security solutions by considering the different security and efficiency requirements of data communications.
Hence, we propose a differentiated security mechanism for protecting infotainment data communication in IoV networks. In particular, we first classify data communication in the IoV network, examine the security focus of each data communication, and then develop a differentiated security architecture to provide security protection on a file-to-file basis. Our architecture leverages Named Data Networking (NDN) so that infotainment files can be efficiently circulated throughout the network where any node can own a copy of the file, thus improving the hit ratio for user file requests. In addition, we propose a time-sensitive Key-Policy Attribute-Based Encryption (KP-ABE) scheme for sharing subscription-based infotainment data. Coupled with our NDN network, any node in the network can own the file, but only those with a valid subscription can decrypt the file, thus achieving the security of infotainment data distribution while optimizing content distribution efficiency. 

\end{abstract}

\section{Introduction}
The Internet-of-Vehicle (IoV) is an integration of Vehicular Ad-hoc Networks (VANET), and the Internet of Things (IoT) \cite{surveyIoV20}. The IoV is a network of vehicles that uses various sensors, software, built-in hardware, and communication technologies to enable information exchange between pedestrians, cars, and urban infrastructure \cite{ANT}. With seamless connectivity, IoVs can perform various functionalities, such as providing real-time navigation guidance and delivering onboard infotainment services.

IoV infotainment systems use touchscreen displays, button panels, and audio/video interfaces in the vehicle to provide information and entertainment. It is connected to onboard units via Control Area Network (CAN) to integrate various functionalities and offer a unified user interface for both entertainment (such as retrieving media content from nearby transportation infrastructure and presenting it, connecting to personal devices to enable onboard voice/video projection), and driver-assistance (such as real-time monitoring of vehicle conditions via installed cameras, receiving real-time traffic status updates, etc.). It also acts as a platform for translating input from the user into messages communicated across the IoV network via built-in Bluetooth, cellular, and Wi-Fi modules. Although there is a unanimous agreement on the importance of protecting traffic communication security \cite{surveyIoV20, v2xsurvey20, mollah2020blockchain, antsurvey}, infotainment systems have not received adequate attention in their communication security. Negligence on the security of infotainment data communication in IoV networks can unintentionally open an easy access point for social engineering attacks, where attackers can influence a person psychologically to act in the interest of the attacker\cite{fi11040089}. For example, the attacker can spread false information about traffic conditions through vehicle-to-vehicle infotainment communication, mislead drivers into crowding a particular highway and interfere with traffic management by reporting false traffic information. Furthermore, there will be distractions for drivers as they attempt to find accurate information after realizing it is incorrect, which could have a potential implication for driving safety. Hence, infotainment data communication security is important to data integrity in IoV networks and road safety.

In the literature on IoV communication and network security, most security solutions aim to provide secure communication via a single mechanism, such as user authentication using strong cryptographic operations and are growingly unfit for increasingly complicated network communication. Although these operations are required to secure crucial traffic control messages on the IoV network, they often have a high computational cost and a substantial overhead for vehicular communication. However, not all messages on the IoV network require this high level of security protection. Systems can be more efficient by reducing the complexity of the security framework and ensuring that important communications are efficiently and adequately protected by tailoring security protections according to the type and nature of data exchanged. Such an adaptation can be termed Quality-of-Security-Service (QoSS).

Similar to Quality-of-Service (QoS), Quality-of-Security-Service (QoSS) is a concept that measures the level of security in a system where there is a variable level of security services and requirements against which the system is evaluated. As a result, not all the information exchanged in the IoV ecosystem will have the same level of QoSS, and not all will require stringent authentication protection. The degree of security factors considered would subsequently be referred to as the system's level of security. By matching security requirements with security mechanisms that offer such capabilities, QoSS can enhance system performance and achieve higher user satisfaction by providing users or network tasks with a range of appropriate security choices \cite{QoSScynthia01}.

In the case of the IoV network, due to the wide range of messages transmitted in IoV communications, different types of data can have different security and efficiency requirements, i.e. different QoSS. Depending on the nature of the infotainment content, i.e., whether the media is private or publicly available, there will be a different set of security and efficiency expectations. Public infotainment data does not require confidentiality protection, while subscription-based infotainment data needs access control based on the user's subscription status. For example, Tesla has preinstalled in-car entertainment applications, such as Netflix and Youtube, in their digital system, and users have to pay a premium subscription fee to access these entertainment applications. For infotainment services on the go, they install WiFi and cellular modems in each Tesla, and users can connect to the car, nearby WiFi networks, or personal/public hotspots, or they can pay a subscription fee and get an LTE connection with the preinstalled SIM card to enjoy infotainment services. Naturally, there will be different security expectations for such subscription services than for unpaid public services. At the same time, the same protocols that safeguard private data exchanges are not suitable for public data because public data are supposed to be accessible by all and have lower criticality for instant and long-term availability. Similarly, the communication protocol for traffic control messages has a different QoSS from public data, with top priority on maintaining the integrity and availability of the information while achieving low latency. Hence, the security design of the IoV communication network can be more effective and efficient if we consider the different QoSS of the communication to the different security mechanisms that we have in place.

This paper aims to provide differentiated security protection for infotainment data communication in IoV networks. The security mechanisms utilised to protect each communication are determined by the type of data exchanged and the security focus of such communication, i.e., the QoSS of each communication. The main contributions of this paper are summarised as follows:
\begin{itemize}
\item We classify data communication in the IoV network into six categories and examine its security focus. By doing so, we analyse the different QoSS of each type of data communication in terms of confidentiality, integrity, short-term availability, and long-term availability.
\item We apply caching with Named Data Networking (NDN) towards infotainment data communications to improve resource distribution efficiency. Since infotainment resources such as music and films are often substantial in size, satisfying every user request by retrieving these files from the server would impose a significant communication overhead. As a result, the combination of caching and NDN network can provide an effective distribution mechanism to circulate these files throughout the network to fulfil every user request at the shortest possible hop distance.

\item We develop a differentiated security architecture for securing infotainment data communication in IoV networks, leveraging the NDN technique and designing security protection on a file-to-file basis, ensuring that various types of data receive appropriate protection and eliminating the wastage of resources as in generic security solutions. By decoupling the resource's identity from the provider's identity, e.g., IP address, the data packets are self-contained and independent of where they can be retrieved and transferred. This design allows infotainment files to be easily circulated throughout the network, where any node can keep a copy of the file, thus improving the hit ratio when file requests are made.

\item We propose a time-sensitive Key-Policy Attribute-Based Encryption (KP-ABE) scheme for sharing subscription-based infotainment data communication where the user can only decrypt files whose validity time falls within their subscription time. When the validity of a user's access policy expires, the user loses the ability to decrypt the files until they resubscribe from the service provider. Coupled with our NDN architecture, any node in the network can own the file, but only those with a valid subscription can decrypt the file.

\item \textcolor{black}{We integrate blockchain technology to manage access to subscription-based infotainment data and avoid revocated users from decrypting subscription infotainment data.}
\end{itemize}

The structure of this paper is in accordance with the following: 
In Section \ref{sec: related}, we present the related work to introduce the context of the paper.
In Section \ref{sec: overview}, we provide an overview of the differentiated security architecture for securing infotainment data communication in the IoV network and explain our rationale and methodology.
In Section \ref{sec: kp-abe}, we propose a time-sensitive KP-ABE scheme for the sharing of subscription-based infotainment data communication and provide our evaluation of this scheme.
In Section \ref{sec: conclusion}, we conclude our work in this paper.

\section{Related Work \label{sec: related}}

\subsection{Internet-of-Vehicles}
Given recent advancements in telecommunication technologies and more powerful computation engines, we expect our vehicles to be more intelligent and capable. IoV has come to light as a promising direction for solving the growing demand for a safer, faster, and more comfortable transportation network in cities. The IoV is a combination of Vehicular Ad-hoc Networks (VANET) and the Internet of Things (IoT). VANET is a network made up of a group of vehicles and roadside units connected by a wireless network. IoV can be seen as a superset of VANET, which has a considerable improvement over VANET. An overview of the difference between VANET and IoV can be found in \cite{SHARMA2019100182}\cite{surveyIoV20}, where the authors conclude that the impediment to the commercialization of VANET is its unreliability and limited scalability to provide seamless connectivity. 

The communication types in IoV include vehicle-to-vehicle (V2V) and vehicle-to-infrastructure (V2I) \cite{ali2019authentication}, while Vehicle-to-everything (V2X) is a combination that maximizes the advantages of both V2V and V2I \cite{barrachina2013v2x}.
To better understand the transformation of vehicle-to-everything (V2X) technology towards IoV, a survey on the challenges and opportunities arising from the evolution of V2X technology to IoV is presented in \cite{v2xsurvey20}, where the authors also proposed an improved cloud-IoV architecture enhanced with Mobile Edge Computing (MEC) as a vision of future IoV technologies. The study of IoV covers a wide range of topics, from resource optimization for communication \cite{cognitivecarrier21} to efficient charging systems utilizing the smart grid to optimize autonomous driving routes.
Besides these topics, security and privacy challenges in IoV have attracted a lot of attention from the community because the normal operation of the IoV system is inseparable from secure, reliable, and efficient network communication and data exchange. 
Attackers may manipulate the real-time traffic control messages or public vehicular traffic data streams, eavesdrop on V2X private information, or intercept subscription-based infotainment data to unauthorized users, which may cause huge economic and social security damages.
Therefore, it is critical to design a secure and efficient data exchange scheme using identity authentication and privacy-preserving techniques in IoV.

\subsection{Caching With Named Data Networking}
The efficiency of infotainment data communication in mobile networks heavily relies on data caching approaches due to the significant size of media content. Traditional communication techniques based on TCP/IP and IP will be less effective in meeting such efficiency criteria in highly dynamic mobile environments such as the IoV systems, especially as the number of devices in the network grows \cite{chen2020caching}. The IoV system often demands a higher transmission rate and more flexibility in data distribution protocols due to the continuous pursuit of better audio and video quality and the limited network bandwidth.

Named Data Networking (NDN) emerges as one of the suitable candidates for resource optimization in such networks, providing both compatibility with existing routing protocols and optimization of communication resource utilization \cite{zhang2014named}.
It performs information-centric networking and can be directly applied to major IP services like Domain Name Service (DNS) and inter-domain routing policies. IP routing protocols like BGP and OSPF can be adapted to NDN with little modifications.
Instead of source and destination addresses, NDN nodes send data packets by their content names. Unlike IP routers, the use of unique content identifiers for communication allows routers to keep track of packet states, allowing them to perform a variety of activities. The data packets are self-contained and independent of where they can be retrieved and transferred. These characteristics enable in-network caching of content for future requests, enhancing content mobility while eliminating the requirement for application-specific middleware. Furthermore, NDN routers allow multi-path forwarding, which means they can route a user request to numerous interfaces simultaneously.

The authors in \cite{yan2014novel} first proposed to leverage NDN in the vehicular information network environment to provide efficient communication and scalable information retrieval services.
Later, many works investigated NDN-based research topics in Vehicular Ad-hoc Networks (VANET) domain.
In \cite{khelifi2019named}, the authors provided a comprehensive review of the state-of-the-art works and challenges on NDN-based VANET.
The system architecture and packet forwarding phase of NDN were explained in their work. The naming schemes, routing and forwarding mechanisms, data transmission, in-network caching, mobility support solutions, security and privacy, and simulation techniques of NDN in VANET have also been surveyed in detail. After all, NDN has shown its ability to provide a safe, efficient, and infotainment-rich driving service for users. Hence, we also leveraged NDN for data dissemination and secure and efficient subscription-based infotainment data communication in IoV.

\subsection{Attribute-based Encryption}
The ability to provide infotainment services in IoV is a major commercialization contributor to the continuous development of IoVs. Besides public infotainment services like radio channels or free media content, a significant portion of infotainment data, such as video and music files, is supported by third-party service providers and is often subscription-based. 
Attribute-based Encryption (ABE) \cite{6883405} is a common strategy for providing access restrictions to support subscriber-only access.
In IoV systems, some research works leveraged the ABE techniques for protecting data security and privacy during the data transmission and caching processes. 
The authors in \cite{zhang2021enabling} used CP-ABE to guarantee IoV data security when outsourcing and sharing IoV data in the cloud and fog. 
CP-ABE is also applied in \cite{taha2019cluster,deng2022secure,zhao2022verifiable} for secure data sharing in VANET.
In addition to data confidentiality, the short-term availability of infotainment resources should also be considered to ensure secure and efficient access control of subscription-based infotainment services in IoVs.
The work in \cite{liu2018time,liu2020new} has proposed a time-based Ciphertext-Policy Attribute-Based Encryption (CP-ABE) scheme by utilizing direct revocation and maintaining a short revocation list by a secret key time validation method. The secret key time validation method can deny access to users whose access has exceeded their subscription time. They have managed to keep their revocation list short, as keys of expired users will be removed from the revocation list. 

While CP-ABE encrypts messages with access policies and distributes user keys to the policies according to user attributes, KP-ABE is the exact opposite. In our work, we proposed a time-sensitive KP-ABE where the AES keys to decrypt files are encrypted with attributes that only privileged users' access policies can match. In this way, a user can have a broad policy tree where different combinations of keys can be used to satisfy and decrypt, allowing greater flexibility in access management. A ``platinum user", for example, should be able to decrypt any content that a ``gold user" can decrypt, assuming ``platinum" is a higher tier subscription class than ``gold". An important advantage of doing so is the relatively small amount of overhead for data protection that is being circulated in the NDN network. Unlike in CP-ABE where a large-sized access policy will be encrypted with the files to cater to potential users of varying status and characteristics, KP-ABE encrypts the files with the few attributes that a privileged user should have.


\subsection{Blockchain}
Blockchain has been widely used in IoV for decentralized, secure, transparent, immutable, or automatic data access control, identity management, and other applications
\cite{mollah2020blockchain}. 
Network nodes can audit the data access control and identity verification to avoid unauthorized access and malicious tampering \cite{guo2018using}.
Blockchain is utilized for secure and trustworthy data sharing in \cite{kang2019toward,khelifi2020blockchain,ren2021novel}, and leveraged for identity management in \cite{tan2019secure,lin2020bcppa,yang2021blockchain,son2022design} for IoV.
In addition, the authors in \cite{shi2020blockchain} applied blockchain to protect multimedia data sharing from being tampered with or forged in IoV.
A blockchain-based certificate revocation approach is proposed in \cite{lei2020blockchain} for reducing the communication overhead and
shortening the user revocation processing time in the intelligent transportation system.
To properly manage access to subscription-based infotainment data of vehicles, we also leverage blockchain in our work to verify the user's pseudo-identity, expected expiration, and transaction timestamps to avoid revocated users from decrypting subscription infotainment data.

\section{Overview of System Architecture\label{sec: overview}}
The ability to provide infotainment services on the go is vital to the adoption of IoV systems. At the same time, the demand for a safer road environment and transportation infrastructure continues to be the main motivation for smart transportation. Furthermore, secure communication with effective and efficient privacy protection and user anonymity becomes a critical requirement due to the dynamic and open environment in which these applications operate. 

Motivated by the above findings, we propose a differentiated security architecture for sharing infotainment data in IoV networks. The goal of this architecture is to provide a means for safe and efficient data exchange for public and subscription-based infotainment data, while private infotainment data should be protected using conventional authentication methods.

  \begin{figure*}[htb]
    \begin{center}
        \includegraphics[width=1\linewidth]{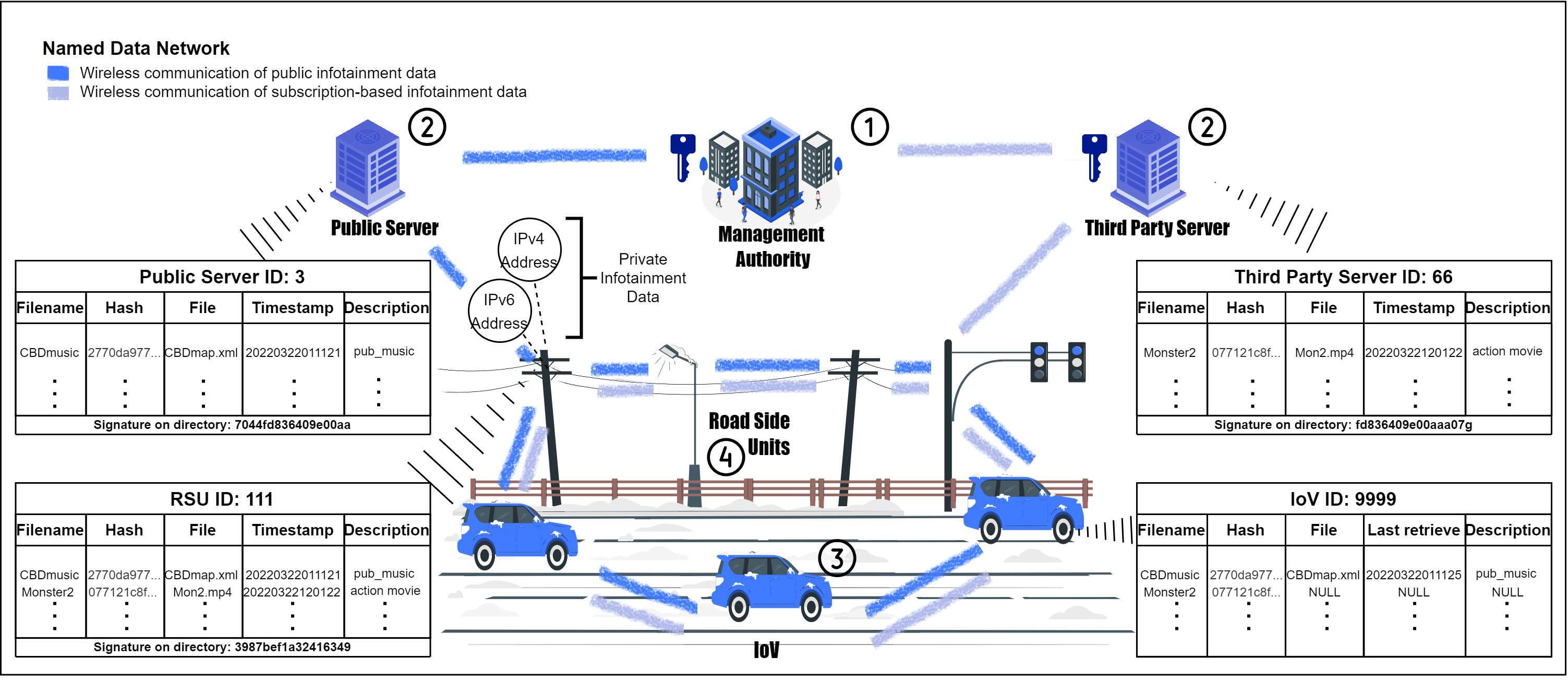}
        \caption{Illustration of the proposed NDN-based communication security architecture.}
        \label{fig:iov_context}
    \end{center}
    \end{figure*}
    
With reference to Figure \ref{fig:iov_context}, a general IoV transport system consists of four main components: (1) management authority, (2) storage servers, (3) IoVs, and (4) road-side units. The management authority (1) is the central agency that monitors the traffic status in real-time and performs traffic control operations to ease road conditions. It is also responsible for disseminating important traffic information to its subsidiary networks, such as wide-area networks consisting of roadside and traffic signalling infrastructure. Data generated here is commonly stored on central or remote cloud servers (2). Due to the high mobility of IoVs, roadside units (4) and neighbouring vehicles (3) frequently serve as data relays to support efficient traffic information communication and to provide a seamless connection to IoVs.

For the remainder of this section, we first present a classification of common data exchange in an IoV network and demonstrate the diverse security and efficiency requirements of each data type, which leads to major design considerations for security schemes in infotainment data sharing in such IoV networks that have yet to be fulfilled in many generic security schemes. Then, we introduce our model and explain how it can satisfy these requirements.

\subsection{Data Classification}
First of all, we classify data exchange in IoV communication and explain the security focus of data communication in each category to better illustrate the need for differentiated security schemes in such IoV networks.
\begin{enumerate}
\item \textbf{Vehicle-to-everything (V2X) private information exchange:} This includes any private or sensitive information about the vehicle or the user which is not meant to be shared with the public, such as electric billing messages, personalized route planning, or messages that can reveal the user's identity. Since these messages can disclose the user's private information, it is important to protect the confidentiality and integrity of these messages. Furthermore, a heterogeneous pool of devices is integrated for communication with the vehicle, ensuring the availability of services is also vital to the IoV environment.

\item \textbf{Traffic control messages:} These are real-time traffic control messages issued, such as traffic light messages and emergency messages (e.g. ambulance warnings and accident warnings). These messages are time-sensitive and have high requirements for message integrity and availability. Any malicious attempt that disrupts these messages can cause life-threatening consequences. Thus, communication protocols that exchange traffic control messages should implement strong security protocols against attacks on the integrity and availability of these messages.

\item \textbf{Public traffic data:} This includes publicly available information on traffic congestion status, geographic maps, infrastructure maintenance warnings, and even gas stations, fire stations, and car repair service stations. Because such information is intended for all road users, there is no requirement for confidentiality protection. On the other hand, the integrity and availability of this information are critical for road users to make informed decisions.

\item \textbf{Publicly accessible infotainment data:} This covers any publicly accessible infotainment data with no access restrictions. This category includes online websites, social media platforms, and content sharing by service providers that are publicly available for everyone to view.

\item \textbf{Subscription-based infotainment data:} This includes infotainment data provided by third-party content providers based on users' subscriptions. For such services, there is a need for authentication of the user's subscription status. Furthermore, quality of service is crucial, where availability and low latency are the top priorities, necessitating the deployment of effective caching methods.

\item \textbf{Private infotainment data:} This contains infotainment data that is restricted and not intended for the general public to view freely. For example, personal social media account content will require user identification before being accessed since such entertainment data can be private and sensitive. As a result, this type of data will need stringent security measures to ensure its confidentiality and integrity. While the availability of these data services is vital for improving customer experiences, it is less important than ensuring the availability of traffic-related information.

\end{enumerate}

\begin{table*}[hbt]
\caption{Comparison among different security focuses for data categories in IoV.}
\centering 
\begin{tabular}{ |p{3cm}|p{3cm}|p{3cm}|p{3.5cm}|p{3.5cm}|  }
\hline\hline
\bf{Categories} & \bf{Confidentiality} & \bf{Integrity} & \bf{Long-term Availability} & \bf{Short-term Availability} \\
\hline\hline

V2X private information exchange & Highly critical & Highly critical & Critical & Critical\\\hline
Traffic control messages & Moderate & Highly critical & Highly critical & Highly critical \\\hline
Public traffic data & Not applicable & Highly critical & Critical & Moderate\\\hline

Publicly accessible infotainment data & Not applicable & Highly Critical & Important for user experience & Important for user experience\\\hline
Subscription-based infotainment data & Confidential against non-subscriber & Highly Critical & Important for user experience & Important for user experience\\\hline

Private infotainment data & Highly Critical & Highly Critical & Important for user experience & Important for user experience\\\hline

\end{tabular}

\vspace{1ex}
{\raggedright Note: We divide the availability into two categories: short-term availability (the availability of the resource when it was first requested and delivered to the intended user) and long-term availability (the availability of the resource after a significant period of time). \par}
\label{tab: dataCategoryRequirement}

\end{table*}

In Table \ref{tab: dataCategoryRequirement}, we analyze the different QoSS of each type of data communication in terms of confidentiality, integrity, long-term availability, and short-term availability for different data categories in IoV. With the finite computational resources in IoV systems, we can prioritize the security needs according to the type of data communicated and draft different security protocols to meet individual QoSS. 

Private data, such as ``V2X private information exchange" and ``private infotainment data," for example, should be protected by stringent authentication mechanisms to preserve its confidentiality and integrity. On the other hand, the communication protocol for ``traffic control messages" has a top priority in maintaining the integrity and availability of the information while achieving low latency. The same protocols that safeguard the aforementioned data types are not suitable for public data, such as public traffic data and public infotainment data, because these public data are supposed to be accessible by all and have lower criticality for instant and long-term availability. Edge nodes in IoV networks are often desired to store this information for easy distribution. At the same time, peer-to-peer sharing will also help speed up the circulation of public information, but the integrity of the information exchanged is a concern. Similar to public information, subscription-based infotainment data provided by third-party service providers are often cached in edge nodes for fast transmission, requiring additional access restriction by subscription status. 

Thus, we can observe different security and efficiency focuses among data exchanges in IoV networks. IoV network communication can be made more efficient if we provide a differentiated security implementation based on the requirements of different data categories.

\subsection{Design Considerations}

There are several considerations that we made in our design of the security architecture for infotainment data sharing:

\paragraph{\textbf{Differentiated security}} Although critical information, such as traffic control messages and peer-to-peer sharing of road conditions, necessitates strict cryptographic safeguards, less sensitive public information, such as public infotainment data, has a comparatively lower demand for QoSS. Hence, we designed a differentiated security architecture where private infotainment communications are still protected using conventional authentication mechanisms. Also, files containing public infotainment data are hashed, and the directory that contains their filenames and hash values is signed using the authority's public key certificate to protect their integrity. At the same time, we proposed a time-sensitive KP-ABE scheme to facilitate the secure distribution of subscription-based infotainment data. 
\paragraph{\textbf{Minimizing IoV network latency}} Since traffic data needs to be communicated in real-time, the IoV network can only tolerate a small level \textcolor{black}{(e.g. 30-40ms)} of network latency for all network communications. However, infotainment data, like video and music files, might be several gigabytes in size. With the high mobility of IoVs, there is a need for strategic information dissemination mechanisms to enable smooth communication in such networks. Thus, we adopt NDN and cache public and subscription-based information at the RSUs to reduce the hop distance between users and data. 
\paragraph{\textbf{Re-usability of encrypted files}} Since infotainment data files are huge, it is preferable to retrieve them from nearby devices to reduce hop distance, network overhead, and network delay. While all road users have access to public information, all encrypted subscription content in the network should be decryptable by all users whose policies meet the decryption conditions. Hence, subscription contents should not be encrypted using user-specific information but rather conform to the general conditions that privileged users have. Thus, we proposed a time-sensitive KP-ABE scheme that allows all users who have matched subscription attributes and whose subscription validity time completely covers the required time validity on files to decrypt the files. This way, the exact copy can circulate throughout the network and be decrypted by any privileged user who requests it, increasing the hit ratio.

\subsection{Proposed Approach for Infotainment Data Exchange}
This approach focuses on the secure and efficient transmission of infotainment data. Private data communication and traffic control messages will be protected by traditional authentication procedures \textcolor{black}{\cite{7869418,9039557}}, while public traffic data can be stored in files and protected in the same manner as public infotainment files are, up to the discretion of system designers.

In our proposed architecture, illustrated in Figure \ref{fig:iov_context}, \textbf{IoV users} are allowed to request files from neighbouring IoV users or any RSU that has the files, whichever is nearest to the user. If a neighbouring IoV user or RSU does not have the required file, the request is forwarded to the server, which is in charge of distributing the requested file. Aside from public information, IoV users can request any file, such as traffic-related information from the traffic authority, on-board infotainment resources from RSUs, Internet content via URLs, or user data from cloud servers. The response is parsed into NDN format and transmitted back to the IoV user by the nearest RSU that receives the request.

\textbf{RSU} stores a directory of resource names and file hashes. This directory is digitally signed by RSU to ensure its integrity. IoV users can obtain an update of the public directory from the RSUs and verify the digital signature of the RSU against the pre-loaded public key certificates of public infrastructures. In addition, RSUs can cache public traffic data, such as road maps and popular public infotainment data, such as public media channels, which are denoted by their resource names. Furthermore, RSU may conduct IPv4/6 operations, such as retrieving content from URLs, caching it, translating it into Named Data Network format and distributing it to requesters.

To improve the user experience, \textbf{third-party infotainment service providers} can choose to cache popular subscription content at RSUs. These files are encrypted using our time-sensitive KP-ABE on third-party servers with desirable user attributes that the service provider has distributed to its subscribed users. More details on the KP-ABE scheme can be found in the next section. \textcolor{black}{The providers can choose to periodically update the stored directories in RSUs via authenticated communication between providers and RSUs. Since both the filename and the hash value of the encrypted file have been communicated to the RSUs, the encrypted file can be transmitted via untrusted channels, and its integrity can still be checked for.}  For the rest of this subsection, we explain our proposed interactions for the exchange of public traffic data, public infotainment data, and subscription-based data.

\subsubsection{\textbf{Exchange of public infotainment data}} Popular public infotainment data is cached in RSUs, and RSUs have a directory of infotainment data resource files, their hash value, last-update-timestamp and content descriptions, along with the digital certificate issued from RSU. This directory is updated with new incoming files from authorized data providers. In this directory, there are two types of files: default public infotainment files and additional public infotainment files. Once the IoV receives the directory, it verifies the digital certificate against the pre-loaded trusted certificates of public infrastructures and starts retrieving the content.

All default public infotainment information files, except those already existing in the system and have not been modified since the last retrieval, will be downloaded automatically to the IoV. In comparison, additional public infotainment files are only distributed when IoV users request them. An IoV in transit can request a new copy of the directory from the RSU or from nearby vehicles, whichever is in close proximity. Users can modify their system settings to always download the desirable media content (the next part of the movie the user is watching) when it is available. Large media files are broken down into smaller files so that the expected transmission time per file is low. Given the short amount of interaction time between fast-moving IoV and nearby RSU or neighbouring IoVs, this will lower the number of incomplete transmissions. These download requests are fulfilled by either the RSU or peers. Once the download for the entire file is complete, the IoV verifies the hash value of the file against the directory entry to check for integrity.

IoV could verify the hash value of the portion they received completely and request other portions from neighbours in proximity.

\subsubsection{\textbf{Exchange of subscription-based data}} Third-party content providers can cache their content in the RSU to improve the user experience or retrieve the content from their server when RSU receives the request. They encrypted the content with the user attributes such as membership status, account validity, and decryptable time period (the time period that the file can be decrypted) using the time-sensitive KP-ABE scheme. The subscription infotainment application (a user copy of the entertainment software installed) on board will have a default content directory that stores the popular content on the platform. For example, ``Monster 2" is popular content among IoV users, and the service provider has included the file name of this movie and the hash value of the encrypted file in the default resource directory, which is pre-installed in the infotainment application and updated on a regular basis.

When IoV user request a particular resource, e.g. ``Badguy", the IoV infotainment application can first search for the corresponding file name and hash value, i.e. ``badguy.mp4" and the hash value of encrypted ``badguy.mp4". If it is not present in the default list, it can request the file name and hash value of the file from the RSU. Once the IoV has the file name and hash value, it can request the file from RSU or neighbouring IoVs, and verify the hash value. Once verified, the IoV infotainment application will decrypt the file using its access policies. 

\begin{figure*}[!hbt]
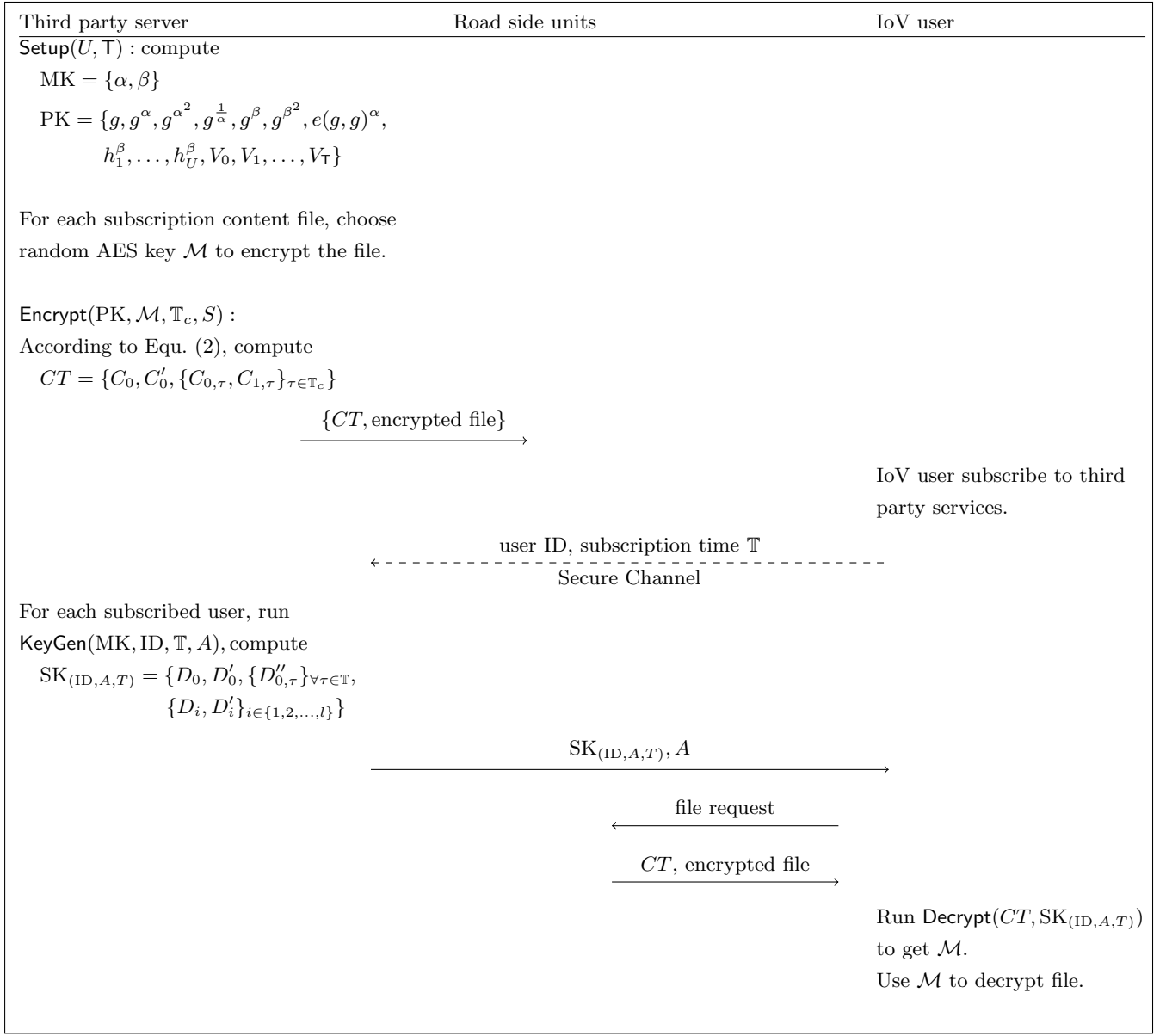

\centering
\fbox{
\procedure{}{
 \text{Third party server} \< \< \text{Road side units} \< \< \text{IoV user}  \\[] [\hline]
 \mathsf{Setup}(U, \mathsf{T}): \text{compute} \< \< \< \< \\
\t \mathrm{MK}=\{\alpha, \beta\} \< \< \< \< \\
\t \mathrm{PK}=\{g, g^{\alpha}, g^{{\alpha}^2}, g^{\frac{1}{\alpha}}, g^{\beta}, g^{\beta^2}, e(g,g)^{\alpha},\< \< \< \< \\
\t\t\t\t h_1^{\beta},\dots,h_U^{\beta}, V_0, V_1, \dots, V_{\mathsf{T}} \} \< \< \< \< \\
\< \< \< \< \< \\
\text{For each subscription content file, choose} \< \< \< \< \\
\text{random AES key $\mathcal{M}$ to encrypt the file.} \< \< \< \< \\
\< \< \< \< \\
\mathsf{Encrypt}(\mathrm{PK}, \mathcal{M}, \mathbb{T}_c, S): \< \< \< \< \\
\text{According to Equ. (\ref{equ:ct}), compute}\< \< \< \< \\
\t CT=\{C_0,C_0',\{C_{0,\tau}, C_{1,\tau}\}_{ \tau\in\mathbb{T}_c}\} \< \< \< \< \\
\< \sendmessage*{->}{top={\{CT, \text{encrypted file}\}}, width=0.5cm} \< \< \< \\
\< \< \< \< \text{IoV user subscribe to third}\\
\< \< \< \< \text{party services.}\\
\< \< \< \sendmessage{<-}{top={user ID, subscription time $\mathbb{T}$}, bottom={Secure Channel}, style={dashed}, length=8cm, width=1cm} \< \\
\text{For each subscribed user, run} \< \< \< \< \\
\mathsf{KeyGen}(\mathrm{MK}, \mathrm{ID}, \mathbb{T}, A), \text{compute} \< \< \< \< \\
\t \mathrm{SK}_{(\mathrm{ID}, A, T)}=\{D_0, D_0', \{D_{0,\tau}''\}_{\forall \tau\in \mathbb{T}},\< \< \< \< \\
\t\t\t\t\t\t\t \{D_i, D_i'\}_{i\in\{1,2,\dots,l\}}\} \< \< \< \< \\
\< \< \< \sendmessage*{->}{top={\mathrm{SK}_{(\mathrm{ID}, A, T)}, A}, length=8cm, width=1cm} \< \\
\< \< \< \sendmessageleft{top={file request}} \< \\
\< \< \< \sendmessageright{top={$CT$, \text{encrypted file}}} \< \\
\< \< \< \< \text{Run}~\mathsf{Decrypt}(CT, \mathrm{SK}_{(\mathrm{ID},A,T)})\\
\< \< \< \< \text{to get}~ \mathcal{M}.\\
\< \< \< \< \text{Use}~\mathcal{M}~\text{to decrypt file.}\\
}}
\caption{Workflow of the proposed time-sensitive KP-ABE scheme for subscription-based infotainment data communication.}
\label{fig:exchanges}
\end{figure*}

The workflow of the proposed time-sensitive KP-ABE scheme for subscription-based infotainment data communication is presented in Figure \ref{fig:exchanges}. The secure channel here is established using user authentication so that users can verify their accounts and make purchases for their subscriptions. Every time a user makes a new subscription, a new pseudo ID is generated based on the user's ID and the current subscription time.

\subsubsection{\textbf{Exchange of private infotainment data}} Private infotainment data can contain sensitive personal information. Hence it should be protected using authentication mechanisms to prevent unauthorized access to such data.

\subsection{Subscribed User Revocation}
A different pseudo-identity is generated by the service provider based on the user's identity every time they purchase their subscription. If a subscribed user decides to cancel their membership before the expiry date, the service provider can cancel the subscription and place the user's current pseudo-identity, expected expiry and transaction timestamp on the permissioned blockchain maintained by the service provider. On a daily basis, the infotainment service agent installed on IoV will query the user's current pseudo-identity to check for revocation details. If it is found, the agent will mark the user as a revoked user and stop serving the user until a new pseudo-identity is issued to the user, which also means the start of a new subscription.

When the expected expiry of the transactions stored on a permissioned blockchain is reached, those transaction records are removed from the blockchain since these pseudo-identities can no longer be used to decrypt any files from this date onwards.

\section{Time-sensitive KP-ABE Scheme \label{sec: kp-abe}}

In this section, we introduce our Key-Policy Attribute-Based Encryption (KP-ABE) scheme for the secure distribution of subscription-based infotainment data in IoV networks.

Inspired by the work \cite{liu2018time, liu2020new}, we use a Hierarchical Identity-based Encryption (HIBE) based approach to control the time validity of the infotainment files. In general, the time periods are represented by a hierarchical tree which has one topmost root node and, at most three-level non-root nodes. Each node in the first level of the tree represents a year, and its child in the second level represents a month in this year. The third level nodes represent the days.
Note that we also adopt the set-cover approach to select the minimum number of nodes to represent all the valid time periods. \textcolor{black}{The use of HIBE in conjunction with the set cover approach can effectively reduce the number of key generations required to represent each time period.} For example, if a user purchases a subscription service from 2022-JUL-01 to 2022-SEP-02, the tree of his valid time periods should contain four nodes, including {2022-JUL, 2022-AUG, 2022-SEP-01, 2022-SEP-02}. Therefore, this user should obtain the corresponding attribute in their access policies for those four nodes by using HIBE in order to decrypt the ciphertext within the valid time periods. For example, the user can only decrypt files whose validity time period falls within his/her subscription time period, meaning files that have a validity time period that is equal to or is a subset of 2022-JUL-01 to 2022-SEP-02. When the validity of a user's access policy expires (after 2022-SEP-02), the user loses the ability to decrypt until they resubscribe from the service provider.

Our scheme consists of a 4-tuple of algorithms, denoted as $(\mathsf{Setup}, \mathsf{KeyGen}, \mathsf{Encrypt}, \mathsf{Decrypt})$, of which the construction details are shown below: 
\begin{itemize}[leftmargin=10pt]
\item $\mathsf{Setup}(U,\mathsf{T})$: it is a setup algorithm that takes the number of attributes $U$ and the depth of the time tree $\mathsf{T}$ as input, outputs the public parameters $\mathrm{PK}$ and a master key $\mathrm{MK}$. In specific, given the depth $\mathsf{T}$, each time period is represented as a z-ary string $\{1,z\}^{\mathsf{T}-1}$, i.e, \{2022, 09, 22\}.
The algorithm chooses a bilinear group $\mathbb{G}_1$ of prime order $p$ with a random generator $g$, and randomly selects $U$ elements from the group, i.e. $h_1, h_2, \dots, h_U \in \mathbb{G}_1$. Besides, it also randomly chooses $\alpha,\beta \in \mathbb{Z}_p$ and $V_0, V_1, \dots, V_{\mathsf{T}} \in\mathbb{G}_1$. 
Then, it outputs
\begin{equation*}
\begin{aligned}
&\mathrm{MK}=\{\alpha, \beta\},\\
\mathrm{PK}=&\{g, g^{\alpha}, g^{{\alpha}^2}, g^{\frac{1}{\alpha}}, g^{\beta}, g^{\beta^2}, e(g,g)^{\alpha}, h_1^{\beta},\dots,h_U^{\beta}, V_0,\\ &V_1, \dots, V_{\mathsf{T}} \}.
\end{aligned}
\end{equation*}

\item $\mathsf{KeyGen}(\mathrm{MK}, \mathrm{ID}, \mathbb{T}, A)$: for a specific user with a pseudo-identity $\mathrm{ID}$ (A different pseudo-identity is generated by the service provider based on the user's identity every time that he/she purchases his/her subscription) and a set-cover of decryptable time periods, denoted as $\mathbb{T}$ that each of the elements in $\mathbb{T}$ can be represented as a $z$-ary representation $\tau=\{\tau_1, \tau_2, \dots, \tau_k\}\in\{1,z\}^k$ where $k<\mathsf{T}$, give the master key $\mathrm{MK}=\{\alpha, \beta\}$ and the LSSS access structure $A=\{M, \rho\}$, where $M$ is an $l\times n$ matrix and $\rho$ is a mapping function that maps each row of $M$ into an attribute. This algorithm outputs a private key $\mathrm{SK}_{(\mathrm{ID}, A, T)}$ for this user according to the following operations. At first, it chooses a random masking vector $\boldsymbol{v}=\{w, y_2,\dots,y_n\}\in \mathbb{Z}_p^n$ to share the encryption exponent $w$. Besides, it computes $\lambda_i = \boldsymbol{v} \cdot M_i$ for $\forall i\in \{1, 2, \dots, l\}$, i.e. $M_i$ is the $i$-th row vector of $M$. Here $\{\lambda_i\}$ are the shares of the secret $w$ according to $M$. Then this algorithm can calculate
\begin{equation}
\label{equ:sk}
\begin{aligned}
&D_0=e(g,g)^{\alpha w},~ D_0'=g^{\frac{w}{\alpha}},~\\ &\left\{D_{0,\tau}''=\left(V_0\prod_{j=1}^kV_j^{\tau_j}\right)^w\right\}_{\forall \tau\in \mathbb{T}},~\\
&D_i=g^{\beta\lambda_i},~
D_i'=\left(gh_{\rho(i)}^{\beta}\right)^{\lambda_1\mathrm{ID}}
\end{aligned}
\end{equation}
and get the private key 
$$\mathrm{SK}_{(\mathrm{ID}, A, T)}=\{D_0, D_0', \{D_{0,\tau}''\}_{\forall \tau\in \mathbb{T}}, \{D_i, D_i'\}_{i\in\{1,2,\dots,l\}}\}$$

\item $\mathsf{Encrypt}(\mathrm{PK}, \mathcal{M}, \mathbb{T}_c, S)$: this is the algorithm that uses the public key $\mathrm{PK}$ generated by the $\mathsf{Setup}$ algorithm to encrypt a plaintext message $\mathcal{M}$\footnote{$\mathcal{M}$ is generally the \textcolor{black}{256 bits} AES key used to encrypt the actual content because the size of the actual content is generally larger than the maximum size of the message that can be encrypted by ABE schemes.} associated with a set of attributes $S$ and a set of decryptable time periods $\mathbb{T}_c$. 
The set $S$ consists of attributes such as film rating and subscription tier (e.g., platinum, gold, and silver). 
The set $\mathbb{T}_c$ consists of some time elements $\tau=\{\tau_1', \tau_2', \dots, \tau_{k_{\tau}}'\}\in\{1,z\}^{k_{\tau}}$ where $k_{\tau}<\mathsf{T}$. The set $\mathbb{T}_c$ is determined by the content provider. For example, if the provider decides that the content is valid for a particular period,  $\mathbb{T}_c$ will cover that period so that only users who subscribed for this period will be able to decrypt.

The algorithm chooses a random $x \in \mathbb{Z}_p$ and for $\forall \tau\in \mathbb{T}_c$, it chooses a random $v_{\tau}\in \mathbb{Z}_p$. It then computes 
\begin{equation}
\label{equ:ct}
\begin{aligned}
&C_0=\mathcal{M}\cdot e(g,g)^{\alpha x},~ 
C_0'=g^{{\alpha}^2x},~
C_{0,\tau}=g^{v_{\tau}},\\
&C_{1,\tau} = g^{\alpha x}g^{{\beta}^2}\left(V_0\prod_{j=1}^{k_y}V_j^{\tau_j'}\right)^{v_{\tau}}
\end{aligned}
\end{equation}
where $k_y=(g^{\beta} h_y^{\beta})^{-1}$ for $y\in S$. Finally, it outputs the ciphertext $CT=\{C_0,C_0',\{C_{0,\tau}, C_{1,\tau}\}_{ \tau\in\mathbb{T}_c}\}$ along with the time periods $\mathbb{T}_c$. 

\item $\mathsf{Decrypt}(CT, \mathrm{SK}_{(\mathrm{ID},A,T)})$: this algorithm takes as input the ciphertext $CT$ and a user's private key $\mathrm{SK}_{(\mathrm{ID},A,T)}$, and outputs $\bot$ if any one of the following situations occurs:
\begin{enumerate}
    \item $S$ does not satisfy the access structure $A=\{M.\rho\}$.
    \item $T$ is not completely covered in $\mathbb{T}_c$, i.e. $\tau_T$ and all its prefixes are not in $\mathbb{T}_c$.
\end{enumerate}
Otherwise, let $I=\{i:\rho(i)\in S\} \subset \{1,2,\dots,l\}$, there exists a set of constants $\{\omega_i\in\mathbb{Z}_p\}_{i\in I}$ satisfying that $\sum_{i\in I}\omega_i \lambda_i=w$, where $\lambda_i$ are valid shares of a secret $w$ according to $M$. Finally, this algorithm can decrypt $CT$ as
$$\frac{C_0\cdot e(D_0'',C_{0,\tau}\cdot e(C_0',D_0'))}{e(C_0',g^{1/\alpha})\cdot\prod_{i\in I}\left(e\left(C_{1,\tau},(D_i')^{\frac{\omega_i}{\mathrm{ID}}}\right)\cdot e(D_i,k_{\rho(i)})^{\omega_i}\right)}.$$
\end{itemize}

The user with the appropriate access policy whose valid time period completely covers the decryption time period of the file can achieve the message decryption by solving this equation above.

The correctness of the scheme is shown below.

\begin{equation*}
\begin{aligned}
&\frac{C_0\cdot e(D_0'',C_{0,\tau}\cdot e(C_0',D_0'))}{e(C_0',g^{1/\alpha})\cdot\prod_{i\in I}\left(e\left(C_{1,\tau},(D_i')^{\frac{\omega_i}{\mathrm{ID}}}\right)\cdot e(D_i,k_{\rho(i)})^{\omega_i}\right)}\\
&=\frac{\mathcal{M}\cdot e(g,g)^{\alpha x}\cdot e\left((V_0\prod_{j=1}^kV_j^{\tau_j})^w, g^{v_\tau}\right)\cdot e(g^{{\alpha}^2x},g^{\frac{w}{\alpha}})}{e(g^{{\alpha}^2x},g^{\frac{1}{\alpha}})\cdot\prod_{i\in I}\left(e\left(g^{\alpha x}g^{{\beta}^2}\left(V_0\prod_{j=1}^{k_y}V_j^{\tau_j}\right)^{v_{\tau}}, \left(gh_{\rho(i)}^{\beta}\right)^{\lambda_1\mathrm{ID}\frac{\omega_i}{\mathrm{ID}}}\right)\cdot e\left(g^{\beta\lambda_i}, (g^{\beta} h_{h(i)}^{\beta})^{-\omega_i}\right)\right)}\\
&=\frac{\mathcal{M}\cdot e(g,g)^{\alpha x}\cdot e\left((V_0\prod_{j=1}^kV_j^{\tau_j})^w, g^{v_\tau}\right)\cdot e(g^{{\alpha}^2x},g^{\frac{w}{\alpha}})}{e(g^{{\alpha}^2x},g^{\frac{1}{\alpha}})\cdot e\left(g^{\alpha x}g^{{\beta}^2}\left(V_0\prod_{j=1}^{k_y}V_j^{\tau_j}\right)^{v_{\tau}}, g^w\right)\cdot e(g^{\beta w},g^{-\beta})}\\
&=\frac{\mathcal{M}\cdot e(g,g)^{\alpha x}\cdot e\left((V_0\prod_{j=1}^kV_j^{\tau_j})^w, g^{v_\tau}\right)\cdot e(g^{{\alpha}^2x},g^{\frac{w}{\alpha}})\cdot e(g,g)^{{\beta}^2 w}}{e(g^{{\alpha}^2x},g^{\frac{1}{\alpha}})\cdot e\left(g^{\alpha x}g^{{\beta}^2}\left(V_0\prod_{j=1}^{k_y}V_j^{\tau_j}\right)^{v_{\tau}}, g^w\right)}\\
&=\frac{\mathcal{M}\cdot e(g,g)^{\alpha x}\cdot e(g^{{\alpha}^2x},g^{\frac{w}{\alpha}})}{e(g^{\alpha x},g^w)\cdot e(g^{{\alpha}^2x},g^{\frac{1}{\alpha}})}\\
&=\mathcal{M}
\end{aligned}
\end{equation*}

\subsection{Model Evaluation}
Similar to the work~\cite{liu2018time, liu2020new} that we built upon, the security of our time-sensitive KP-ABE scheme is based on the modified decisional q-parallel-BDHE assumption.
That is, as long as this assumption holds, no adversary can selectively break our time-sensitive KP-ABE scheme.

Consider a modified decisional q-parallel-BDHE game between a challenger $C$ and the adversary $H$. In this game, the adversary aims to decrypt a challenge ciphertext without having the right private key.

\begin{itemize}
     \item Initialization: the challenger $C$ accepts an attribute set $S*$ and a decryptable time period $\mathbb{T}_c*$ from adversary $H$ 
     \item Setup: $C$ runs the \emph{Setup} algorithm and generates the public parameters $PK$
     \item Phase 1. $H$ requests for several private keys corresponding to the identity $ID$, access structure $A*$, and the range of validity time periods $\mathbb{T}$ such that every returned private key satisfies at least one of the following conditions:
     \begin{enumerate}
         \item $S*$ does not satisfy the access structure $A*$
         \item $\mathbb{T}_c*$  is not completely covered in $\mathbb{T}$
     \end{enumerate}
     
     \item Challenge. $H$ submits two equal length message $m_0$ and $m_1$. $C$ flips a fair coin $\beta \in \{0,1\}$ and encrypts $m_\beta$ under the attribute set $S*$ and the time $\mathbb{T}_c*$. The resulting ciphertext $CT*$ is given to $H$.
     \item Phase 2. this phase is completely the same as Phase 1.
     \item Guess. $A$ outputs a guess $\beta'$ of $\beta$.
\end{itemize}

In this game, adversary $H$ is said to have a negligible advantage if the likelihood of $H$ making a correct guess, i.e. $\beta'$ = $\beta$, is barely more than or equal to 0.5. 
We can argue that our time-sensitive KP-ABE scheme is secure if all polynomial-time adversaries\footnote{They cannot break this game in polynomial time\textcolor{black}{\cite{liu2018time}.}} have at most a negligible advantage in the above game. The proof of the adversary having at most a negligible advantage in such a modified decisional q-parallel-BDHE game would be very similar to the ones given in both ~\cite{liu2018time, liu2020new}. This is because our construction is the same as theirs, except that in ~\cite{liu2018time, liu2020new}, attribute set is used in \emph{KeyGen} algorithm and access structure is used in \emph{Encrypt} algorithm whereas this is reversed in our scheme. However, this does not impact the security model. 

We also compare the efficiency of our scheme with other revocable ABE schemes. Table~\ref{tab:eff} shows the comparison. The first row presents a revocable KPABE scheme, the second and third rows present revocable CPABE schemes, and the last row presents our scheme. We can observe that in terms of the size of the master public key ($PK$), ours is comparable to ~\cite{liu2018time,liu2020new}. 
In our scheme, the size of the private key ($SK$) is linearly dependent on the size of the set of decryptable time periods determined by the content provider and hence our scheme may be inferior to other schemes if the content provider determines that content is to be circulated for a very long period. However, due to the volatile nature of media content, the content providers may not set the period for too long in practice. \textcolor{black}{For example, if the content provider determines that a content is to be circulated for a very long period like 10 years, the maximum size would be $60$ (days) $+ 22$ (months) $+ 10 (years) - 1$.}
In our context, the size of Ciphertext is important so that RSUs and IoVs can store and distribute them efficiently. In our scheme, the Ciphertext size is linearly dependent on $\mathbb{T}_c$, which is the decryptable period for the particular subscribed user. In comparison, in other schemes~\cite{datta2016adaptively,liu2018time}, the Ciphertext size is dependent on the size of the revocation list and the number of user attributes. Hence, the schemes may be comparable in the size of Ciphertext.
On the other hand, our scheme is clearly the most efficient in decryption time. Given that short-term availability requirement is critical in our context, it is of most important to us.

\begin{table}[htb]
	\centering
	\caption{Efficiency comparison}
	\label{tab:eff}
	\begin{tabular}{lcccc}
		\hline
		Scheme & PK size & SK size & Ciphertext size & Decryption time (\#pairing)\\ \hline
		\cite{datta2016adaptively} & $111\mathbb{G}_1 + \mathbb{G}_2$ & $(5+16l+16(log^{2}N+logN))\mathbb{G}_1$ & $(16S + 64R - 27)\mathbb{G}_1 + \mathbb{G}_2$ & $16S+37$ \\
		
		\cite{liu2018time} & $(U + R + \mathsf{T} + 3)\mathbb{G}_1 + \mathbb{G}_2$ & $(S + Z + R + 1)\mathbb{G}_1$ & $(l + 3)\mathbb{G}_1 + \mathbb{G}_2$ & $2I + 4$ \\
		
		\cite{liu2020new} & $(U + \mathsf{T} + 3)\mathbb{G}_1 + \mathbb{G}_2$ & $(S + Z + 1)\mathbb{G}_1$ & $(2lR' + 1)\mathbb{G}_1$ + $\mathbb{G}_2$ & $2IR' + 2$  \\
		
	    Ours & $(U + \mathsf{T} + 7)\mathbb{G}_1 + \mathbb{G}_2$ & $(2l +  |\mathbb{T}| + 1)\mathbb{G}_1 + \mathbb{G}_2 $ & $(2|\mathbb{T}_c| + 1)\mathbb{G}_1 + \mathbb{G}_2$ & $2I + 3$ \\ \hline
	\end{tabular}
    \vspace{1ex}
    
    {\raggedright $U$: maximum number of attributes in the system, $N$: maximum number of users in the system, $\mathsf{T}$: depth of time tree, $S$: number of attributes of the user, $I$: number of attributes used in the decryption, $l$: number of rows of the access structure matrix, $Z$: best case $Z=2$ and worst case $Z=(\mathsf{T}(\mathsf{T}+2)/2)$, $R$: maximum number of revoked users, $R'$: length of the revocation list, $|\mathbb{T}|$: number of elements in the set-cover $|\mathbb{T}|$, $|\mathbb{T}_c|$: number of elements in $\mathbb{T}_c$ (best case $|\mathbb{T}|$ and $|\mathbb{T}_c| = 1$ and worst case $|\mathbb{T}|$ and $|\mathbb{T}_c| = 60$ (days) $+ 22$ (months) $+ ($number of decryptable years$ - 1)$) \par}

\end{table}

\section{Conclusion \label{sec: conclusion}}
In conclusion, communication security for IoV infotainment systems is essential and different types of infotainment data can have different security and efficiency requirements. Systems can be more efficient by reducing the complexity of the security framework and ensuring that important communications are adequately protected by tailoring security protections according to the type and nature of data exchanged. To achieve this, we propose a differentiated security architecture for secure and efficient infotainment data communication in IoV networks, leveraging NDN and designing security protection on a file-to-file basis, ensuring that various types of data receive appropriate protection and eliminating the wastage of resources in generic security solutions. In particular, our time-sensitive KP-ABE scheme allows content providers to set time validity on their content and protect their copyrights. In future work, we plan to investigate incentive mechanisms that encourage IoV users under our system to conduct peer-to-peer infotainment resource sharing, which is a crucial feasibility concern for improving the efficiency of resource sharing and reducing communication overheads for file retrievals.

\bibliographystyle{splncs04}
\bibliography{references}

\clearpage

\appendix
\section{Summary of math notation and symbols}
\begin{table}[hbt]
\centering
\caption{A summary of math notation and symbols.}
\renewcommand{\arraystretch}{1.1}
\begin{tabular}{p{2.5cm}<{\centering}|l}
\hline
Symbol & Description \\ \hline
$U$             & the number of attributes \\ \hline
$\mathsf{T}$    & the depth of the time tree \\ \hline  
$\mathrm{MK}$          & the master key \\ \hline
$\mathrm{PK}$          & public parameters \\ \hline
$\mathbb{G}_1$ & a bilinear group of prime order $p$  \\ \hline
$g$              & a generator of $\mathbb{G}_1$  \\ \hline
$h_1, h_2 \dots h_U$   & random elements chosen from $\mathbb{G}_1$ \\ \hline
$V_1, V_2,\dots,V_\mathsf{T}$   & random elements chosen from $\mathbb{G}_1$ \\ \hline
$\alpha$, $\beta$      & random numbers chosen from $\mathbb{Z}_p$ \\ \hline
$\mathrm{ID}$ & a user's pseudo-identity  \\ \hline
$\mathbb{T}$  & a set-cover of a user's decryptable time periods \\ \hline
$\tau$              & a z-ary representation of a time element \\ \hline
$A$           & a LSSS access structure  \\ \hline
$M$              & an $l\times n$ matrix  \\ \hline
$\rho$              & a mapping function \\ \hline
$\boldsymbol{v}$ & a random masking vector in $\mathcal{Z}_p^n$ \\ \hline
$\omega$              & an encryption exponent  \\ \hline
$\lambda_i (i=1,2,\dots, l)$ & the shares of $\omega$ \\ \hline
$\mathrm{SK}$ & a private key of a user \\ \hline
$\mathcal{M}$ & a plaintext message \\ \hline
$\mathbb{T}_c$   & a set of decryptable time periods of a message \\ \hline
$S$              & a set of attributes of the message  \\ \hline
$CT$          & a ciphertext \\ \hline

\end{tabular}
\label{tab:my-table}
\end{table}

\end{document}